\documentclass[12pt, english]{article}
\usepackage[a4paper, total={16cm, 22cm}]{geometry}
\usepackage{hyphenat}
\usepackage{graphicx} 
\usepackage{indentfirst}
\usepackage[affil-it]{authblk}

\title{Bulgarian national input to the European Strategy for Particle Physics}

\author[2]{L. Anguelova}
\author[1]{M. Bogomilov}
\author[1]{M. Chizhov}
\author[2]{A. Demerdzhiev}
\author[1]{K. Dimitrova}
\author[1]{K. Gladnishki}
\author[2]{R. Hadjiiska}
\author[2]{P. Iaydjiev} 
\author[1]{S. Ilieva}
\author[1]{D. Kocheva}
\author[1]{D. Kolev}
\author[1]{K. Kostova}
\author[1]{V. Kozhuharov}
\author[1]{F. Kunis}
\author[1]{L. Litov}
\author[2]{M. Makariev}
\author[2]{G. Maneva}
\author[1]{D. Mihaylov}
\author[1]{M. Naydenov}
\author[1]{B. Pavlov}
\author[1]{P. Petkov}
\author[1]{G. Rainovski}
\author[3,2]{M. Shopova}
\author[1]{R. Simeonov}
\author[2]{G. Soultanov}
\author[2]{P. Temnikov}
\author[2]{D. Tonev}
\author[1]{R. Tsenov}
\author[1]{G. Vankova-Kirilova}
\author[1]{V. Verguilov}

\affil[1]{{Faculty of Physics, Sofia University ''St. Kl. Ohridski'', 5 J. Bourchier Blvd., 1164 Sofia, Bulgaria}}
\affil[2]{Inst. for Nucl. Research and Nucl. Energy, Bulgarian Academy of Sciences, Tsarigradsko shose,  BG-1784 Sofia, Bulgaria}
\affil[3]{Faculty of Physics and Technology,
University of Plovdiv ''Paisii Hilendarski'', 24 Tsar Assen Str., Plovdiv, Bulgaria}

\date{April 2025}

\begin{document}

\maketitle
\begin{abstract}
The present document summarizes the view and the vision of the Bulgarian subatomic physics scientific community 
for the development of the field of subatomic physics 
from the Bulgarian national perspective. 
It outlines the present activities and the  
strengths and weaknesses of the research field, 
together with the interests of the community for 
the future development of the technological and
scientific landscape. 

\end{abstract}

\newpage
\section{Introduction}

After the discovery of the Higgs
boson at LHC, the very first fundamental scalar,
the Standard model was assumed to be complete
and expectations were for rather quick
new findings extending our knowledge of the microworld.
However, the last decade in subatomic physics
acted as a sobering period among the community,
with no groundbreaking discoveries
which could answer the long-standing fundamental questions
for the nature of dark matter, the
origin of matter-antimatter asymmetry in the
universe, the unification of the fundamental forces and etc.
In addition, the understanding  of several phenomena plays 
an important role for our description of Nature:
\begin{itemize}
    \item the mechanism for neutrino mass generation,
    \item the transition from perturbative QCD to
how the hadrons are formed from quarks and gluons and what their masses are
    \item understanding the structure of the nucleons including charge and spin distribution;
and chiral symmetry restoration 
    \item the contribution of 
ultra high energy cosmic rays
to the abundance of heavy elements 
\end{itemize}
and are far from being complete. 

On the other hand, significant progress has been achieved in several
directions, some of them highlighting the role of precision and low background measurements, which related Particle Physics to other research areas:
\begin{itemize}
 \item The precision in various flavour related measurements, 
 e.g. $K^+ \to \pi^+\nu\bar{\nu}$ and 
 the anomalous magnetic moment of the muon, increased significantly.

 \item The understanding of the strong interactions at different stages,
 perturbative with QGP and low energy interaction description
 between the final state hadrons, achieved through correlation functions,
 reflected to our knowledge of the equation of state of compact astrophysical objects.

 \item Understanding the mixing in the leptonic sector through the measurement 
 of the PMNS matrix parameters.
 
\end{itemize}

Going beyond particle physics, important progress in the 
last decade was seen in the 
multimessenger astronomy, namely the discovery 
and the utilization of the Gravitational Waves as the new probes
of the Universe.
Various activities, 
such as complex data analysis and anomalies identification 
are seen as common problems and may profit from 
closer interaction between the communities. 

The present document outlines the commitment of the
Bulgarian subatomic physics community  
in the present and future initiatives and
underlines the national interest 
and prospects to strengthen the 
capabilities for such research in the short and 
mid-term scale.

\section{Participation in particle and nuclear physics experiments and facilities}

\subsection{LHC experiments}
\subsubsection{CMS}
Bulgarian scientists have a long-standing and active presence in the CMS experiment since 1999, contributing significantly to its construction, operation, and scientific output. Over the years, Bulgarian researchers and engineers have played an important role in the development and maintenance of key detector subsystems, notably the Hadron Calorimeter (HCAL), the Resistive Plate Chambers (RPC), and the Gas Electron Multiplier (GEM) systems. These contributions have been instrumental in ensuring the high performance and reliability of the CMS detector throughout its operational lifetime.
Beyond technical work, the Bulgarian teams have also taken on several leadership and coordination roles within the CMS Collaboration. Members of the Bulgarian institutions have served in management positions, contributed to task forces and working groups, and received CMS awards recognizing their dedication and impact. These roles underscore the visibility and influence of the Bulgarian scientific community within one of the world’s leading high-energy physics experiments.
As the CMS experiment prepares for the challenging operational conditions of the High-Luminosity Large Hadron Collider (HL-LHC), a comprehensive upgrade of the detector is underway. This effort aims to ensure that CMS remains capable of delivering high-quality data in the high-radiation and high-occupancy environment expected during HL-LHC operations. The Bulgarian scientific community fully supports this critical upgrade program, acknowledging its importance in enabling CMS to fully exploit the physics potential of the HL-LHC over the next two decades.
At present, the Bulgarian team remains deeply engaged in multiple facets of the CMS program. This includes detectors R\&D, ongoing operation and maintenance of existing systems, and active contributions to the development of advanced software tools, including Monte Carlo software and machine learning algorithms. 
Bulgarian researchers also participate in a broad range of physics analyses, with particular emphasis on B physics, 
flavor studies, and vector-like quarks (VLQ) searches, 
which are among the flagship programs of the CMS physics portfolio.   
In addition, interest is expressed in  broad resonance searches,
which further enhances the CMS and LHC physics reach and could lead 
to new physics discovery.
The continuity and expansion of the involvement of Bulgarian scientists is vital—not only for the success of CMS and the HL-LHC, but also for the development of Bulgarian science, technology, and human capital. The Bulgarian community strongly supports the sustained engagement of its teams in CMS, and seeks to further strengthen their contributions.

\subsubsection{ALICE}
Bulgaria became an official member of the ALICE collaboration in 2021 and 
since then the Bulgarian team plays active role in the
ALICE upgrade program, contributing to the 
design, beam tests, and construction of the 
FoCal detector. 
Initially, the Bulgarian interest was strongly related to
the ALICE forward physics program including 
gluon saturation studies and 
beyond the Standard Model UPC events. 
Given the expertise in data analysis, software, 
scintillation detectors and calorimeters, 
the natural choice was the hadron part of the FoCal detector. 
A significant effort is also dedicated to the 
determination of the proper type of photosensor, 
with a continuous campaign for SiPM irradiation 
and performance degradation determination.
In addition, the team plans to support the integration of the detector’s reconstruction algorithms into the O2 framework, ensuring compatibility and efficient data processing within the system. 

Recently, the Bulgarian team was extended with 
core ALICE members, deeply involved in femtoscopy studies. 
The physics interests range from correlation functions and 
interaction potential determination 
to particle source description and collectivity 
studies in small systems. 
The Bulgarian community has a strong interest in the interdisciplinary opportunities between femtoscopy and astrophysics. In particular, a better understanding of the strong force enables the construction of more realistic nuclear equations of state, which are especially relevant for describing the composition of compact astrophysical objects such as neutron stars.
More exotic scenarios—such as the possible presence of axions in neutron stars—can also be indirectly explored through femtoscopy, by imposing stringent theoretical constraints on neutron star properties using nuclear physics alone, as well as through studies of the pion decay constant. This line of research benefits from the Bulgarian community’s established expertise in axion physics. Further point of interest for the Bulgarian community is the recently demonstrated connection between femtoscopic source analyses and coalescence studies. This will allow to study the formation of bound systems—particularly light (anti-)nuclei, ultimately leading to a better understanding of the production properties of (anti)nuclei in our universe, providing valuable input for indirect dark matter searches in space
Another key area of interest for the Bulgarian community is the recently demonstrated connection between femtoscopic source analyses and coalescence studies. This connection enables the study of the formation of bound systems—particularly light (anti-)nuclei—and ultimately contributes to a deeper understanding of their production in the universe. Such insights can provide valuable input for indirect dark matter searches conducted in space.

Precision measurements in UPC, 
namely the anomalous magnetic moment 
of the $\tau$-lepton and searches 
for axion like particles through their di-photon decays 
allow access to fundamental electroweak and precision 
physics, 
with possible sensitivity  to Dark Matter explanation
and several beyond the Standard
Model scenarios. 
In addition, many interesting phenomena could be 
addressed through the study of the electromagnetic radiation
from the QGP, which can be addressed by electromagnetic
component of the Forward Calorimeter (after LS3) and 
further by the ALICE 3 detector. 
In this regard, Bulgarian community supports strong 
commitment to the whole ALICE physics program, 
seeing it also as a bridge 
between particle,  nuclear physics and  astrophysics.

\subsection{Neutrino physics}
The Bulgarian HEP community strongly supports the development of neutrino physics in Europe, building on European experience and recent developments at other global scientific centres in Japan and USA. 
The European Spallation Source Neutrino Super Beam (ESSnuSB) project could be the next-to-next flagship European experiment for long-baseline neutrino research. 
This cutting-edge initiative aims to precisely measure CP violation at the second neutrino oscillation maximum. 
By doing so, ESSnuSB will be able to determine the $\delta_{CP}$ phase with an unprecedented precision of less than 8 degrees for any value of the true $\delta_{CP}$, 
thereby shedding light on the origin of matter-antimatter asymmetry in the universe. 
The experiment will utilize the European Spallation Source (ESS) linac in Lund, 
Sweden as a proton driver to generate neutrino beam. 
By leveraging existing infrastructure and reserving 5 MW of average power for the neutrino program, this setup will produce the world's most intense neutrino beam. 
This powerful beam will enable measurements that cover both the first and second neutrino oscillation maxima. To achieve this, far detectors with a fiducial mass of 540 kilotonnes (kt) will be located in an existing mine in Zinkgruvan, Sweden approximately 360 kilometres away from the source.

In addition, the ESSnuSB project aims at measurements of neutrino cross-sections, study of neutrinos created in the atmosphere, in the sun and in supernova explosions and search for nucleon decay.

Unlike ESSnuSB, which operates at energies below 1 GeV and focuses on producing neutrinos other than tau neutrinos, the SND@LHC experiment at CERN pursues different objectives. All three neutrino flavours are accessible, allowing for investigations of flavour universality in the neutrino sector and precise measurements of cross-sections as a function of energy. The measurements span an energy range from 350 GeV to several TeV, effectively bridging the gap between low-energy neutrinos produced by conventional beamlines and ultra-high-energy neutrinos originating from cosmic rays or astrophysical sources. SND@LHC plans to continue data collection at the High-Luminosity LHC (HL-LHC). To accommodate the increased luminosity an upgraded detector will be constructed. This possibly could also lead to the first direct observation of tau antineutrinos.

\subsection{ISOLDE}

ISOLDE is a unique and world-leading facility for radioactive beam science that uses of GeV protons delivered by the PS Booster to produce a wide range of isotopes used in a diverse science program. The program has continuing science themes such as measurements of ground-state properties of nuclei, radioactive decay, nuclear reactions with post-accelerated beams, and the use of radioisotopes as probes for condensed matter physics. There are also emerging key science topics including searches for beyond Standard Model phenomena in radioactive molecules and $\beta$ decays, probing the distribution of nuclear magnetization, applications of high sensitivity NMR with polarized ions in biology and engineering, anti-protons and exotic nuclei, transfer-induced fission and characterization of novel materials important for sustainable technologies.

During CERN LS3, the facility will be improved to enable use of protons of higher energy and intensity improving the production yields of exotic nuclei which, along with other infrastructure improvements, will enhance both the capability and capacity of the facility. The opportunities provided by these enhancements will be exploited during the subsequent running periods. The Collaboration is developing a plan for further facility improvements during LS4, as well as for new instruments to enable novel measurements. On a longer-time scale, ISOLDE users proposed ambitious plans for an additional experiment hall and new target stations. The evolution of both the facility infrastructure and the associated instrumentation will secure a leading position for ISOLDE into the longer term. As the roadmap for the post-LHC future is developed, a strategy will be required to secure long-term opportunities for continuing world-leading nuclear-physics programmes at ISOLDE, along with the other facilities using the proton injector chain that are unique to CERN.



\subsection{Searching for new light particles}
Being one of the proponents of the PADME experiment at
LNF INFN, the Bulgarian physics community plays a 
key role in the search of new particles in the MeV range 
at positron-on-target annihilation technique. 
Several subsystems were developed and 
were made operational with significant Bulgarian 
contribution. 
The active contribution at all stages of the experiment
made it possible recently to release the 
results of the search for the hypothetical X17 particle
in the mass range around $M_X \simeq 17$~MeV. 
The result, although not conclusive, 
strikes with the precision and detailed treatment 
of systematics and paves the road to further 
higher statistics study of the existence of X17. 

For the Bulgarian particle physics community 
PADME became also  an educational facility to train 
the new generation of particle physicists in experimental techniques
ranging from detector design and construction, data acquisition and 
detector control systems, 
commissioning and calibration, 
contemporary statistical data analysis. 
More than 10 bachelor and master degree thesis 
were successfully defended 
and  
six PhD students got trained or continue their training on 
the PADME experiment. 

This further underlines the importance of 
the existence of small-scale facilities where 
many of the complex and nowadays hidden 
components of a high energy physics experiment 
can be followed in details and understood down to the 
very last bit of technique and technology. 

In this regard, the Bulgarian HEP community
strongly supports and will insist for 
the expansion of such facilities
around Europe, 
especially when the focus is put on local 
initiative in small EU countries, to help to preserve
and further develop the potential of any 
local HEP community. 
{\bf Small scale experiments probing 
non-investigated areas of parameter space 
for a certain physics process, 
or probing an already investigated areas
but with unique techniques
should be supported at all levels} to 
boost the local strength of the particle physics 
communities, 
and to pave the road to new technologies and new ideas for
enlightening fundamental and applied research.

\subsection{Astroparticle Physics}
Astroparticle Physics is a field of particle physics that explores both the micro- and macroscopic worlds. Bulgarian participation in astroparticle physics experiments began in 2005 with joining MAGIC. MAGIC is a system of two 17-meter-diameter Imaging Atmospheric Cherenkov Telescopes (IACTs), located on the Canary Island of La Palma. These telescopes observe gamma rays emitted by cosmic objects in the very high energy range—from 30 GeV to 100 TeV. IACTs are ground-based telescopes that detect Cherenkov light emitted by relativistic particles that exceed the speed of light in air. These particles are produced by interactions between incoming high-energy gamma rays and the Earth's atmosphere. Such telescopes have opened a new window in the electromagnetic spectrum for astronomical observation. Over more than 20 years of operation, MAGIC has discovered many new sources in this energy range, including the first observation of a gamma-ray burst reaching TeV energies. In addition to gamma-ray observations, it has been demonstrated that MAGIC can also measure the fluxes of charged particles and nuclei. The Bulgarian group has played a major role in this area by developing software based on deep neural networks for data analysis.

The next generation of ground-based IACTs, the Cherenkov Telescope Array Observatory (CTAO), is currently under construction and commissioning. It will consist of more than 60 telescopes of three different sizes: Large-Sized Telescopes (LSTs) with a diameter of 23 meters, Medium-Sized Telescopes (MSTs) with a diameter of 11.5 meters, and Small-Sized Telescopes (SSTs) with a diameter of 4.3 meters. Each type is optimized for a specific energy range, and collectively they will enable gamma-ray detection from 20 GeV to 300 TeV. The telescopes will be located at two sites: La Palma (Northern Hemisphere) and Chile (Southern Hemisphere). Bulgaria participates in the LST collaboration at La Palma, where four LSTs are planned for deployment. Currently, one LST is operational, and three others are in an advanced stage of construction. The Bulgarian group is contributing to the optical system of the LSTs.

\section{Prospects and interest of Bulgarian experimental research community}

\subsection{High luminosity LHC}

Given the present participation in the LHC physics program, 
HL-LHC is of primary priority for a large part of the Bulgarian HEP 
community.  
However, its priority should not be defined by the number of running years, 
but considering the desired precision for the measurement of the 
quantities of interest. 
If such precision is reached before achieving 
3000 $fb^{-1}$ and before 2040, 
then HL-LHC should be stopped 
to allow for the
construction 
and operation
of a new facility.

\subsection{Future prospects}
Higgs, being the only fundamental scalar within the Standard model 
has to be studied in details 
and its properties determined with the utmost precision, 
including triple and quartic coupling, couplings to fundamental fermions, etc. 
The most prominent machine for such precision studies
is a lepton collider. 
However, the particular optimal choice for such a facility 
is still unclear to the Bulgarian community 
and no preference exists towards 
circular (e.g. FCC-ee, muon collider) or linear 
lepton collider.
For example, the advantage of FCC-ee is the possible future upgrade to 
an hadron machine at the energy frontierr.
The advantage of a linear collider is the new technology, 
which naturally attracts interest and new manpower and will build 
an ecosystem, around which a sustainable community growth 
could be expected. 

Due to the impossibility to choose among the 
individual options due to lack of
accelerator and infrastructural expertise, 
Bulgarian participation in future initiatives 
will be through development of new detector technologies for
any 
future large-scale facility. 
The detector technologies are 
quasi-invariant with respect to the chosen future facility 
and thus the current Bulgarian support lies 
in the 
research of new detectors and new sensors 
for precision measurements in particle physics. 
This includes the 
development of novel gaseous detectors, 
where recent results have shown that 
substantial performance improvement
can be achieved through a novel design, 
proposed by members of Bulgarian HEP community.

\subsection{Possible steps for fostering a strong particle physics research}

Given the present status of the existing resources in 
terms of manpower and facilities in the country, 
Bulgaria currently acts as a donor 
of experts mainly for CERN, but also for other 
laboratories around the world. 
In addition, most of the experts 
are being trained at the necessary level
through mid- and long-term 
stays at the CERN or other institutions, 
making the cost for a single expert person extremely high. 
And rarely, some of the researchers 
return back to Bulgaria for training the next generation of scientists.
This situation is seen as not sustainable, 
it leads to a drain of the local pool of experts 
and will lead to the impossibility to attract new young 
people to the field in general.
And after a certain time it will reflect the 
capacity of the country to participate actively in the 
fundamental particle physics research,
both CERN or worldwide. 

A possible solution to this problem is seen 
in the construction of a local infrastructure 
with a specific role in the research programme, 
managed by CERN. The infrastructure could be 
an irradiation facility to test various detector 
and electronics components, a low-energy (up to 100 MeV) electron 
accelerator, or any other local facility. 
CERN could 
{\bf push research in the member countries by developing research centers in the form of satellite laboratories. These laboratories may have a very specific tasks as detector research and development or detector performance assessment. }




    

\section{Bulgarian young researchers vision to the development of subatomic physics}
The young Bulgarian researchers community in high energy physics consists of PhD students, postdoctoral researchers and individuals with permanent academic positions. The interests of the community are spread across a wide range of topics with young researchers being part of a number of collaborations and experiments both at CERN as well as in many other laboratories and facilities around Europe. 

A major concern, shared throughout the community, is the uncertainty about the future, both in terms of opportunities for securing positions and generally about the development of the projects they are working on and the availability of funding. For that reason, the whole junior researchers community in Bulgaria is united in the opinion that efforts towards opening new positions, including permanent, scholarships and other means of support for young scientists should be of high priority for the whole high energy physics community. Providing such support would bring benefits to young researchers coming from countries with less funding opportunities, open new career paths and stimulate more young people to pursue academic careers. Moreover, it would also open the possibility for scientists from all across the world to join the research facilities in such countries, something which would lead to strengthening the diversity and expertise of the groups and making strong international connections.

A number of early-career researchers in Bulgaria are part of groups, working on the LHC experiments. There is a strong interest in developing the High-Luminosity LHC and many young researchers are working on detector development and characterization which would be used in the project. Both the CMS and ALICE groups have the desire to continue their work throughout the HL-LHC phase. For that reason, the community is entirely united around the idea that HL-LHC should be prioritized, with the main goal being achieving the planned physics goals rather than running for a fixed number of years.


The opinions about which future project should be supported are divided, with a slight preference towards the FCC. Since the traditions in the Bulgarian community are in detector studies, most of the ECRs express a desire to work on the detector part of whichever project proceeds to be developed. However, some people think that effort should be put into developing an accelerator physics group, a task, which would benefit from the support for attracting experts in the field. 


There is an undivided opinion that any current and future projects go hand-in-hand with developing the computational means and researching new analysis techniques. The whole Bulgarian ECR community is very positive towards using machine learning techniques with the presiding position being that effort should be concentrated towards developing physics-informed and explainable AI methods. Contributing to the understanding of how the used techniques actually work would make ML aided analysis not only a fast but also a reliable method for data handling.

\section{Development of theoretical and computational framework}

To fully exploit the results from the present and future particle 
physics experiments, 
the necessary theoretical framework has to be improved in 
several directions:
\begin{itemize}
    \item Understanding the dynamics of strong interactions towards hadron formation 

    \item Improving the theoretical tools (like holographic methods, the use of symmetries and conserved quantities etc.) for the study of non-perturbative physics
    
    \item Understanding the low-energy effective actions of UV-complete fundamental descriptions (like string theory), in order to extract predictions for axion-like particles and their interactions 
    
    \item Studying dark matter models, based on light weakly interacting particles, and deriving distinctions between this type of models and alternatives like new very heavy particles, modifications of gravity on large scales and other possibilities
    
    \item Exploring the theoretical and observational consequences of primordial black holes as a (significant) component of dark matter
    
    \item Investigating compact objects (like black holes, neutron stars and white dwarfs) in extended theories of gravity, in order to extract possible observational features indicating necessity for modifications of General Relativity
  
    \item Developing further the theoretical framework for quantum computing methods, aimed at processing and analyzing large amounts of experimental or observational data 
    
\end{itemize}

\section{Computing}


The Bulgarian high-energy physics community would like to explore access to national and pan-European supercomputing infrastructures -- such as EuroHPC -- as well as our national supercomputer centers, 
with the intention of enhancing key activities in LHC research and detector development. 
By leveraging high-throughput and high-performance computing resources, we hope to improve reconstruction, 
calibration and physics analysis workflows, 
and to integrate machine-learning algorithms into data-analysis pipelines for tasks such as event classification, 
anomaly detection and rapid parameter estimation.

We intend to investigate the use of GPU- and many-core architectures for 
Geant4-based Monte Carlo simulations of both existing and novel detector geometries, 
potentially shortening prototyping cycles for sensor designs and readout concepts. 
We would also like to employ large-scale computing farms to support extensive parameter scans and the training, 
validation and deployment of ML-driven algorithms for real-time data filtering in next-generation tracking systems. 
Through such collaboration, 
we hope to better prepare for the increased data volumes 
expected in the HL-LHC era and to foster cross-disciplinary expertise at the intersection of particle physics, 
computer science and data science -- ultimately strengthening Bulgaria’s contribution to the emerging European HPC and accelerator-based research ecosystem.

\section{Conclusions}

Developing new technologies which could serve
as the future probes of the microworld boosting our fundamental knowledge,
following the example with the measurement
of the gravitational waves and the 
emergence of multimessenger astronomy.


It is necessary for the particle physics community to understand
that nowadays particle physics competes at equal ground with
other science fields for the crown of
being the most appealing to young researchers,
including quantum technology, quantum informatics, 
artificial intelligence, 
neuroscience etc. 

We, as a sub-atomic physics community in Bulgaria, 
consider that quick and focused actions are needed
among the European HEP community and strategy steak holders
at various levels not to lose the 
attractiveness and the leading role of Europe 
and European scientists in the subatomic research field. 
While in the past CERN acted as a such centralized facility
it might be necessary to consider 
decentralization of certain CERN activities mainly 
in countries with possible risk of loss of expertise and 
scientific potential.
CERN, being the only really global subatomic physics hub, 
is also seen as the only institution that can 
reverse such a process and play a global role in
the training of the researchers of tomorrow. 
An active position and actions in 
operating small- to mid- scale facilities of 
CERN in other countries is seen as a possible and highly desirable 
option.

\end{document}